\RequirePackage{fix-cm}
\documentclass[twocolumn]{svjour3}          
\smartqed  
\usepackage{graphicx}
\usepackage{float}
 \usepackage{mathptmx}      
\usepackage{nicefrac}
\usepackage{hyperref}
\usepackage{amsmath}

\usepackage{subfig}

\newcommand{\cd}{c^{\dagger}}
\begin{document}

\title{Comparison Between the $f$-Electron and Conduction-Electron Density of States in the Falicov-Kimball Model at Low Temperature
}


\author{R.~D.~Nesselrodt         \and
J.~Canfield
       \and
         J.~K.~Freericks  
}


\institute{R.~D.~Nesselrodt, J.~Canfield and J.~K.~Freericks \at
              Department of Physics, Georgetown University\\
              37th and O Sts. NW, Washington, D.C. 20057 USA\\
              Tel.: 202-687-5984\\
              \email{rdn11@georgetown.edu}           
}

\date{Received: date / Accepted: date}

\maketitle

\begin{abstract}
The spinless Falicov-Kimball model is one of the simplest models of many-body physics. While the conduction-electron density of states is temperature independent in the normal state, the $f$-electron density of states is strongly temperature dependent---it has an orthogonality catastrophe singularity in the metallic phase and is gapped in the insulating phase. The question we address here is whether the spectral gap is the same for both electron species as $T\to 0$. We find strong evidence to indicate that the answer is affirmative.
\end{abstract}

\section{Introduction}
\label{intro}
\indent The Falicov-Kimball model \cite{originalFK} is perhaps the simplest solid-state model for describing strongly correlated electron systems. The model possesses a Mott-insulator transition, a charge-density-wave ordered phase and can be solved exactly in the limit of infinite dimensions \cite{Brandt}. Initially the model was employed to understand phase transitions in transition-metal oxides~\cite{originalFK}, and has since been used to study a number of strongly correlated systems. Its solution via dynamical mean-field theory has also been reviewed~\cite{jimFK}.

Recently, the Falicov-Kimball model has been used to investigate core-level X-ray photoemission spectroscopy \cite{jimXPS1} \cite{jimXPS2}. It has been known for some time that the localized $f$-electron Green's function in the metallic phase of this model is related to the x-ray edge singularity problem \cite{XRAYEDGE1,XRAYEDGE2} (we only reference papers immediately relevant to this work here; a more complete history of X-ray edge and the Falicov-Kimball model appears elsewhere~\cite{jimXPS2}). In particular, at low temperature, the density of states in the metal displays a power-law divergence with an interaction-dependent power law.

The $f$-electrons, unlike the conduction electrons, possess a non-trivial and highly temperature dependent spectral function. This has been studied previously using the numerical renormalization group \cite{NRG} illustrating the power law divergence of the $f$-electron spectrum at $\omega=0$ in the metallic phase. Here we seek to understand the gapped behavior of the $f$-electron spectral function in the Mott-insulating regime, where NRG approaches are known to have accuracy issues. Particularly, we ask if the $f$-electron gap approaches the (temperature-independent) conduction-electron gap in the limit $T\to 0$ in the insulating phase. This question may seem like it should have an obvious answer, because the $f$-electron dynamics are inherited through their interaction with the conduction electrons. But at nonzero temperature, we clearly see $f$-electron spectral weight within the gap, so the answer to the question is far from obvious.

\section{Formalism}
The spinless Falicov-Kimball model describes two electron species, conduction ($c$) and localized ($f$) electrons, which interact via a local Couloumb repulsion $U$ when occupying the same lattice site $i$. We represent the creation (destruction) of a conduction electron at the site $i$ by the second quantized operator $c^{\dagger}_i$ ($c_i^{\phantom\dagger}$) and of a localized electron by $f^{\dagger}_i$ ($f_i^{\phantom\dagger}$). Assuming a common chemical potential $\mu$ for the two species, the spinless Falicov-Kimball Hamiltonian is given by
\begin{equation}
    \mathcal{H}_{FK} = \sum_{ij}\left (-t_{ij}-\mu\delta_{ij}\right )\cd_ic_j^{\phantom\dagger} +E_f \sum_i f^{\dagger}_i f_i^{\phantom\dagger} + U\sum_if^{\dagger}_if_i^{\phantom\dagger}\cd_ic_i^{\phantom\dagger}
\end{equation}
where $E_f$ is the $f$-electron site energy, and $t_{ij}$ is the hopping matrix, with $t_{ij}=t=t_{ji}$ for nearest-neighbor sites $i$ and $j$ with the hopping integral $t$ a real constant. We work at half filling ($\mu = U/2$ and $E_f=-U/2$) on the infinite-coordination Bethe lattice in equilibrium.\\

Scaling the hopping energy $t$ with coordination number $Z$ as $t = t^{*}/\sqrt{Z}$ then in the limit $Z \to \infty$ we obtain the noninteracting density of states on the Bethe lattice, given by
\begin{equation}
     \rho_{Z\to\infty}(\epsilon) = \frac{\sqrt{4t^{*2}-\epsilon^2}}{2\pi t^{*2}}
\end{equation}
with $t^{*} = 1$ our energy unit and we restrict to $|\epsilon|\le 2t^*$. 

The dynamical mean-field theory (DMFT) approach solves the many-body problem by mapping the lattice problem onto an impurity problem because the self-energy has no momentum dependence. A self-consistent iterative approach is employed to determine the Green's functions, with details given in~\cite{jimFK}. We do not need the full conduction-electron solution here, instead,
we require only the bare time-ordered Green's function for the conduction electrons $G_0(t)$ given by
\begin{equation}
    G_0(t) =-\frac{i}{\pi}\int_{-\infty}^{\infty}d\omega\ \text{Im}\left [\mathcal{G}_0(\omega)\right ] e^{-i\omega t}[f(\omega)-\theta(t)]
\end{equation}
with $f(\omega)$ the Fermi-Dirac distribution, 
\begin{equation}
    \mathcal{G}_{0}(\omega) = \frac{1}{\omega+i\delta +\mu- \lambda(\omega+i\delta)}
\end{equation}
the effective medium, and the dynamical mean-field $\lambda(\omega)$ obtained from the DMFT algorithm
\cite{jimFK}. Here $\theta(t)$ is the Heaviside unit-step function.

The conduction electrons behave like noninteracting particles, in that they have a temperature-independent density of states, but also like interacting electrons, since they have a Mott metal-insulator transition at $U=2$.
\begin{figure}[htb]
\centerline{
  \includegraphics[width=1.5in]{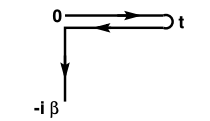}}
\caption{The Kadanoff-Baym-Keldysh contour, which starts at time 0, moves along the real time axis to time $t$, back along the real axis to time 0, then proceeds down the imaginary time axis a distance $-\beta=-1/T$.}
\label{fig:kbk}       
\end{figure}

We define the greater and lesser $f$-electron Green's functions by 
\begin{eqnarray}
    G_f^>(t,t') &=& -i \langle f(t)f^{\dagger}(t') \rangle,\\
G_f^<(t,t') &=&i \langle f^{\dagger}(t') f(t)\rangle,
\end{eqnarray}
 where the angle brackets denote the thermal average, $\langle \ldots\rangle = {\text{Tr}(\exp(-\beta \mathcal{H}_{FK})(...))}/{\text{Tr}\exp(-\beta{\mathcal H}_{FK})} $ The creation and annihilation operators are in the Heisenberg representation. These Green's functions can be determined by selecting $t$, $t'$ on certain branches of the Kadanoff-Baym-Keldysh contour (Fig. \ref{fig:kbk}) and using the contour-ordered Green's function, 
 \begin{equation}
     G_f^c(t_c,t_c') = -i \langle {\mathcal T}_{c}\big(f(t_c)f^{\dagger}(t_c')\big)\rangle 
 \end{equation}
 where $t_c$, $t_c'$ are two times on the contour (Fig. \ref{fig:kbk}). ${\mathcal T}_c$ is the contour time-ordering operator, arranging the operators such that $t_c'$ lies before $t_c$ on the contour. For example, if we pick $t_c'$ on the upper real-time branch of the contour and $t_c$ on the lower real-time branch, we recover $G_f^>(t,t')$ from the contour-ordered Green's function. 
 \\
 
It has been shown in Ref.~\cite{main1}, that in equilibrium the greater Green's functions for the $f$-electrons take the form of a Toeplitz determinant of a continuous matrix operator over \emph{only the positive time branch of the 
contour} 
\begin{equation}\label{det}
    G_f^>(t) = -iw_0e^{-i(E_f-\mu)t}\text{Det}_{[0,t]}\big|\delta(t_1-t_2)-UG_0(t_1-t_2)\big|
\end{equation}
where $w_0$ is the average density of sites without an $f$-electron ($w_0 = \frac{1}{2}$ at half filling), and $G_0(t)$ is the bare time-ordered Green's function determined from the dynamical mean-field $\lambda(\omega)$. The symbols $t_1$ and $t_2$ denote the matrix indices of the continuous matrix operator for which we evaluate the determinant; note that both times must fall within the interval $[0,t]$. (There is a similar expression for the lesser Green's function.) To approximate this continuous matrix operator, we discretize it to a conventional matrix and calculate the determinant for three different discretization time steps $\Delta t$. We then perform a second-order Lagrange interpolation to extrapolate to the limit $\Delta t\to 0$. These numerical results are checked for accuracy against known spectral moments of the Green's functions \cite{SpecMom}.

In the limit of large times, an exact analytic formula for the $f$-electron Green's function in equilibrium can be obtained using a factorization technique from complex analysis described by McCoy and Wu \cite{Mccoy} and called the Wiener-Hopf sum approach. It relies on a result for infinite-sized determinants of Toeplitz matrices called Szego's theorem. Further finite-time approximations can be made to improve the short-time agreement of this asymptotically exact result with the determinant calculation given by Eq.~(\ref{det}) \cite{main1}, with slightly different formulas for the case when the interaction energy $U$ lies above or below  $U_{c}$, a critical interaction strength $U_{c} = \sqrt{2}$ on the Bethe lattice where the complex function $C(\omega) = 1-UG_0(\omega)$ goes from no winding around the origin for $U<U_{c}$ to winding once in the clockwise direction (Ind$\,C$=-1) for $U\geq U_{c}$; note that this critical $U_c$ is smaller than the critical $U$ for the Mott transition. In the regime with $U>U_c$, the analytic result for the Toeplitz determinant in Eq.~(\ref{det}) is 
\begin{eqnarray} 
\label{WH}
    {\rm Det}_{[0,t]} &=& \exp \bigg[\frac{t}{2\pi}\int_{-\infty}^{\infty}\,d\omega\ \ln{\bar{C}(\omega)} + \int_0^{\infty}\,dt'\ t'\bar{g}(t')\bar{g}(-t')\bigg]\nonumber\\
    &\times&\frac{\Delta t}{2\pi}\int_{-\frac{\pi}{\Delta t}}^{\frac{\pi}{\Delta t}} \,d\omega'\ e^{i\omega' t}\frac{\bar{P}(-\omega')}{\bar{Q}(\omega')}
\end{eqnarray}
where $\bar{C}(\omega) = \exp[i\omega\Delta t][1-UG_0(\omega)]$, $\,\bar{g}(t) = \int_{-\infty}^{\infty}\,d\omega$\, \\ $\times\exp[-i\omega t]\ln(\bar{C}(\omega))/2\pi$, and $\bar{P}$   $(\bar{Q})$ are integrals over all positive (negative) time of $\bar{g}(t)$ and satisfy 
\begin{equation}
\bar{C}(\omega) = \frac{1}{\bar{P}(\omega)\bar{Q}(-\omega)}.
\end{equation}
This expression represents a significant reduction in computational complexity, and allows us to probe a much wider parameter space when directly calculating the discretized determinant is not possible. Our approach will be to calculate the determinant directly for short times and use the analytic expression for long times, allowing us to obtain the spectral function of the $f$-electrons down to temperatures significantly lower than previous calculations. We need to patch the two solutions together smoothly, as described below. We examine the behavior of the $f$-electrons as they approach their $T=0$ limit in both the metallic phase near the Mott-insulator transition $U=2$ and in the Mott insulating regime.  

Finally, we define the local density of states of the $f$-electrons, $A_f(\omega)$. At half-filling, there exists a particle-hole symmetry in our system [$A_f(\omega) = A_f(-\omega)$], and consequently the full $f$-electron density of states can be expressed as a Fourier transform of Im $G^>(t)$ alone \cite{Brandt1992},
\begin{equation}
        A_f(\omega) = -\frac{2}{\pi}\int_0^{\infty} \,dt\ \cos(\omega t)\text{Im }G^>(t).
 \end{equation}

\label{sec:1}

\section{Numerics}

We examine the temperature-dependent dynamics of the $f$-electron spectral function in the metallic regime near the Mott transition as well as in weakly and strongly correlated insulating phases above the Mott transition, which occurs at $U=2$. The spectral function is known to have a power-law like divergence in the metallic phase that disappears as we move into the Mott phase \cite{NRG}. In the Mott insulator, the $f$-electron density of states develops a gap with decreasing temperature \cite{main2}. We explore the character of this gap to see if it is the same as in the temperature-independent conduction electron density of states.

\begin{figure}[htb]
\centering{
  \includegraphics[width=0.49\textwidth]{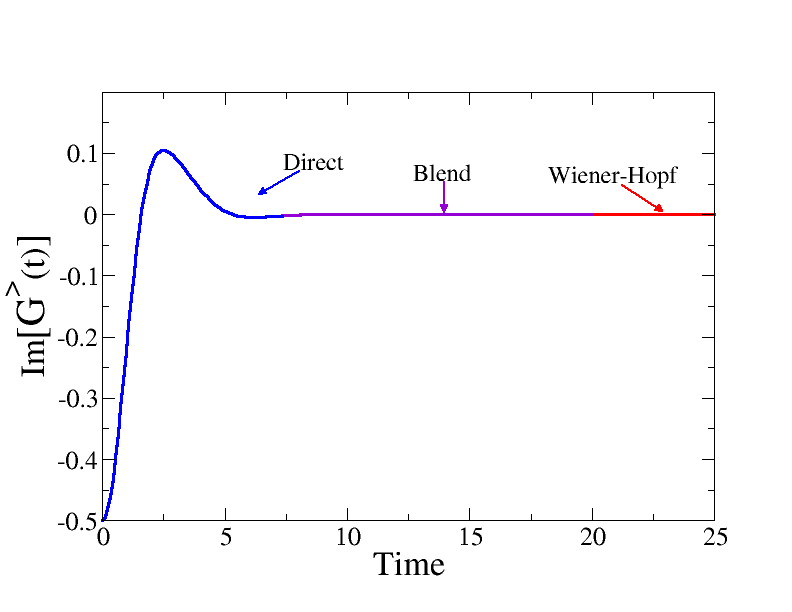}}
\caption{Time-domain plot of Im~$G^>(t)$ for high temperature ($T=1$) in the insulating phase $U\geq 2$. Here the analytic formula works well for all times, and the blending between the two approaches is smooth.}
\label{fig:blend1}       
\end{figure}

\begin{figure}[htb]
\centering{
  \includegraphics[width=0.5\textwidth]{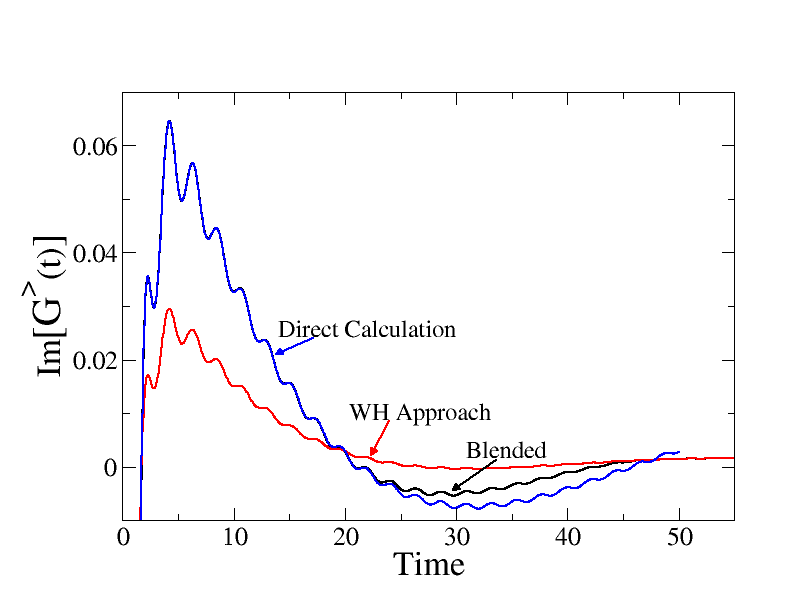}}
\caption{Time-domain plot of Im~$G^>(t)$ for low temperature $T<0.05$ in the insulating phase $U\geq 2$. Here the direct determinant calculation has not reached its asymptotic limit at the maximum time allowed by our computational resources. Note that both Green's functions oscillate with the same frequency, but these oscillations are damped much more quickly when using Eq.~(\ref{WH}). The appearance of high-frequency oscillations at lower temperatures are characteristic of the $f$-electron's low-temperature dynamics.}
\label{fig:blend2}       
\end{figure}

To utilize the Weiner-Hopf technique in Eq.~(\ref{WH}), we perform the discretized matrix calculation of Eq.~(\ref{det}) out to the longest time computationally feasible for three different time steps $\Delta t$ in the ratio $1:2:4$. Next we employ a second-order Lagrange extrapolation to take the limit $\Delta t\to 0$. We use Eq.~(\ref{WH}) to calculate the determinant out to even longer times, and use a weighted blending of the two functions over a time range where the analytic approximation is roughly parallel to the discretized determinant results. This procedure works exceptionally well for high temperature, as shown in Fig.~\ref{fig:blend1}, but as we decrease the temperature, the determinant calculation requires significantly smaller time steps and takes longer to reach the "long-time" regime where our analytic result holds (see Fig. \ref{fig:blend2}). This effect limits our low-temperature calculations, and along with truncating the calculation at a finite time, leads to numerical artifacts near $\omega = 0$ (see Figs. \ref{fig:2DOS}, \ref{fig:25DOS}, \ref{fig:3DOS}).
\begin{figure}[htb]
\centering
  {\includegraphics[width=0.49\textwidth]{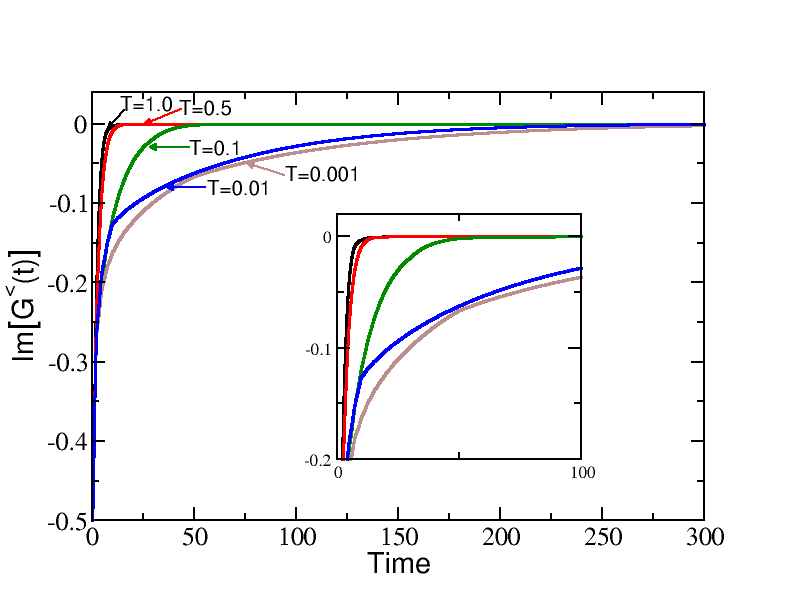}}

\caption{$G^>(t)$ vs.~$t$ for $U=1.5$. Here we have no zero crossing of the time axis, corresponding to the metallic phase of the model. The inset shows the shorter-time behavior.}
\label{fig:15time}       
\end{figure}

\begin{figure}[htb]
\centering{
  \includegraphics[width=0.49\textwidth]{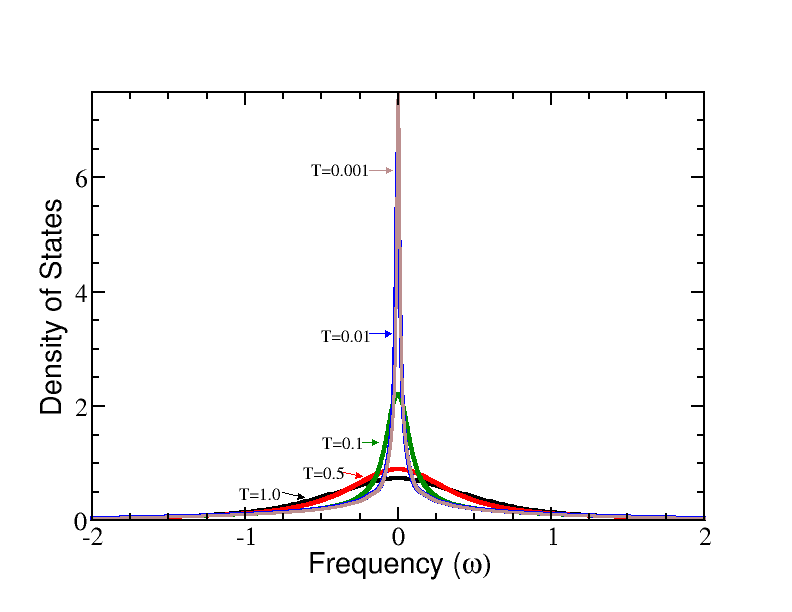}}

\caption{Spectral function $A_f(\omega)$ vs.~$\omega$. Here we see the well documented \cite{NRG} power law divergence of the orthogonality catastrophe set in as $T\to 0$.}
\label{fig:15w}       
\end{figure}

\section{Results}

In the metallic regime, for $U=1.5$ (Fig. \ref{fig:15w}) we see evidence of a power-law divergence as the temperature approaches zero in agreement with \cite{NRG}. In the time domain, we observe a delayed decay towards zero with decreasing temperature. Note there is no zero crossing in Im~$G^>(t)$ (Fig. \ref{fig:15time}), indicating we are still in the metallic phase.

\begin{figure}[htb]
    \centering{
    \includegraphics[width=0.5\textwidth]{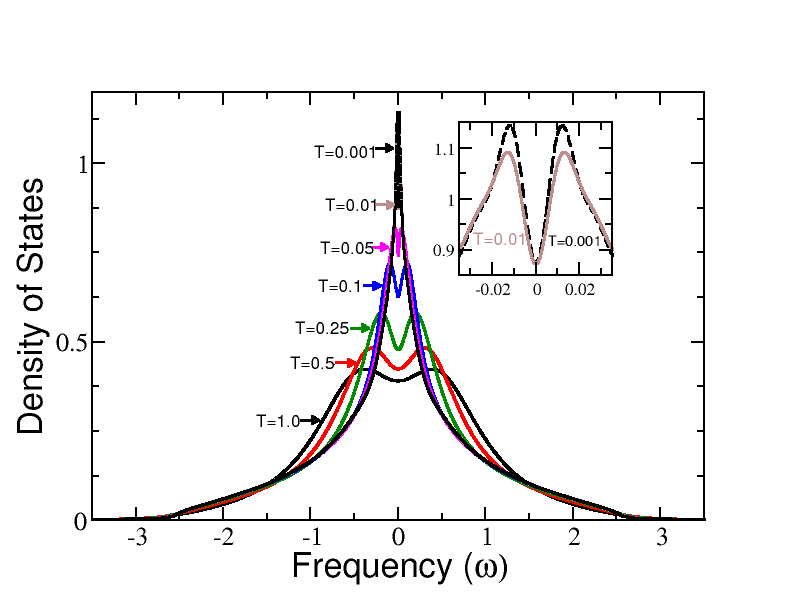}}
    \caption{$A_f(\omega)$ vs.~$\omega$ for $U=1.9$. We see power-law-like behavior, with a central kink that sharpens with decreasing temperature.}
    \label{fig:19}
\end{figure}

For $U=1.9$, closer to the Mott transition, we still see the power-law like increase in the spectral function (Fig. \ref{fig:19}), but with the development of a kink in the center of our density of states, representing a precursor to the insulating Mott phase. 

\begin{figure}[htb]
\centering{
    \includegraphics[width=0.5\textwidth]{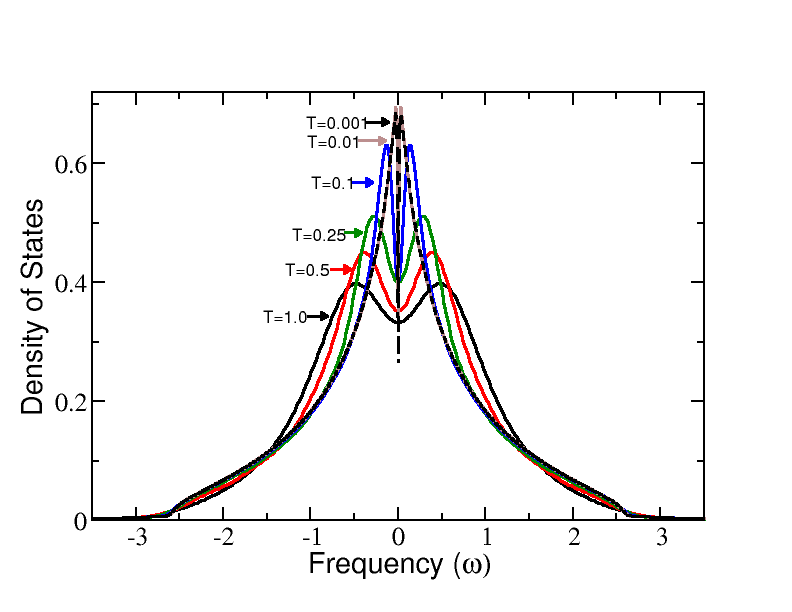}}
    \caption{$A_f(\omega)$ vs $\omega$ at the Mott transition, $U=2$. We see a gradual evolution whereby spectral weight is transferred from higher frequency states to states near $\omega=0$.}
    \label{fig:2DOS}
\end{figure}

At the Mott transition, $U=2$ (Fig. \ref{fig:2DOS}), the $f$-electrons partially fill the insulating gap at high temperature, with a kinked density of states that does not reach zero at the minimum temperature we were able to reach ($T=0.001$). We expect that the density of states should touch zero precisely at $T=0$. This is in contrast with the conduction electrons, whose density of states touches zero at $U=2$ for all temperatures.  

\begin{figure}[htb]
\centering{
    \includegraphics[width=0.49\textwidth]{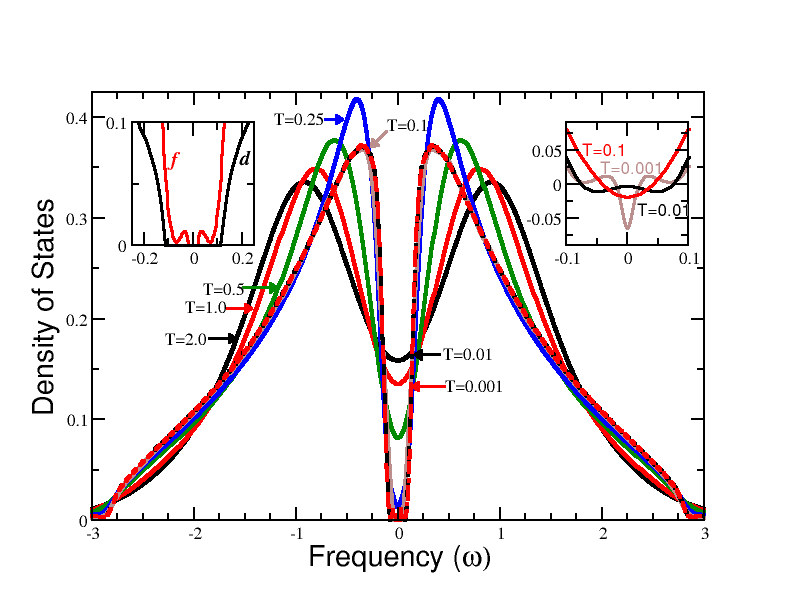}}
    \caption{Temperature evolution of $A_f(\omega)$ vs.~$\omega$ at $U=2.5$ and (inset) the conduction electron density of states over two different ranges of frequency. We see that the gap in the lowest temperature $f$-electron density of states seems to be approaching the gap width of the conduction electron density of states.}
    \label{fig:25DOS}
\end{figure}

For $U=2.5$ (Fig. \ref{fig:2DOS}), we see a similar transfer of weight from high to low frequencies as described in \cite{main2}. Here we see the spectral function becomes gapped around $T=0.1$, but continues to change shape down to $T=0.001$, below which the gap is frozen into place. Notice in the right inset of Fig.~\ref{fig:25DOS} the small region where the density of states becomes slightly negative. This is an artifact of not properly capturing the long-time behavior, likely due to the finite time truncation of some long period oscillations in $G^>(t)$ and possibly the blending of the numerical and asymptotic results.

\begin{figure}[htb]
\centering{
    \includegraphics[width=0.49\textwidth]{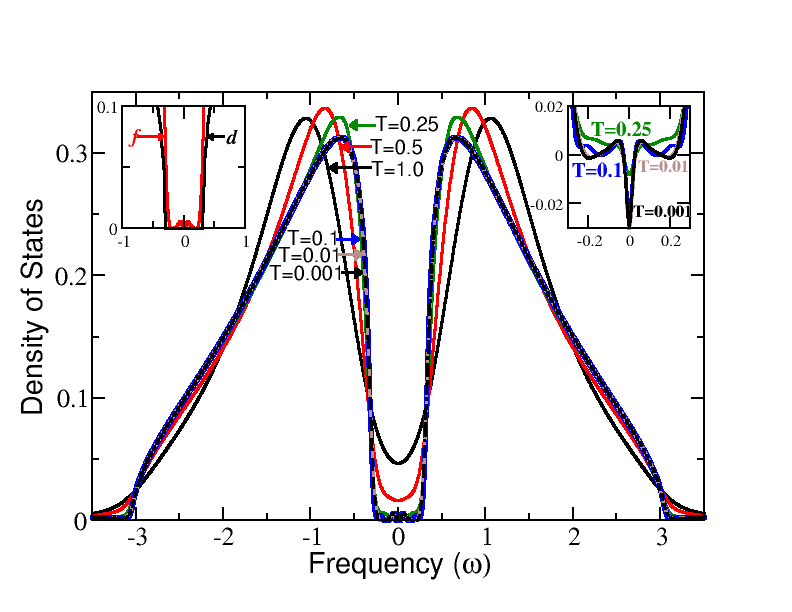}}
    \caption{Temperature evolution of $A_f(\omega)$ vs.~$\omega$ at $U=3$; (inset) the conduction-electron density of states over two different frequency ranges. We again see that the lowest temperature $f$-electron density of states seems to be approaching the width of the conduction electron density of states.}
    \label{fig:3DOS}
\end{figure}

Finally we examine the strongly correlated insulator with $U=3$ (Fig. \ref{fig:3DOS}). We see a similar temperature evolution to $U=2$ and $U=2.5$. In this case, the system becomes gapped around $T=0.25$, higher  than $U=2.5$. Below this temperature $A_f(\omega)$ changes more slowly as the low-temperature behavior is frozen in. We observe a general trend of the gap appearing and freezing into place at a higher temperature for more strongly correlated materials. Again we see the irregular behavior near $\omega = 0$ from the finite time truncation. More interesting, however, is the comparison with the conduction-electron density of states. We see from the left inset of Fig.~\ref{fig:3DOS} that at low temperature the width of the gap is nearly indistinguishable between the two different electrons---this strongly suggests that the $T=0$ gap of the $f$-electrons is the same as the conduction-electron gap. 

\section{Conclusion}

We extensively studied the properties of the $f$-electron spectra of the Falicov-Kimball model in a number of interesting cases: just below the Mott transition, in the strongly correlated regime, and at temperatures approaching the $T=0$ limit. By using the Weiner-Hopf technique, we were able to examine these more complex situations by obtaining an analytic expression for the long-time behavior which is computationally much more efficient than the direct determinant calculation, because the matrices grow in size with increasing time. Due to the rich temperature-dependent dynamics of the $f$-electrons, we asked if the spectral gap in the $f$-spectrum approaches that of the (temperature-independent) conduction electrons at $T=0$. We found strong evidence that this is the case. This technique could be pushed further by making higher-order finite-time corrections to the asymptotic formulas given in \cite{main1} and by carrying the calculations out to longer times.



%

\begin{acknowledgements}
We would like to thank A.~M.~Shvaika for valuable conversations and suggestions. This work was supported by
the Department of Energy, Office of Basic Energy Sciences, Division of Materials Sciences and Engineering under Contract No.~DE-SC0019126.
J.~K.~F. was also supported by the McDevitt bequest at Georgetown.
\end{acknowledgements}

%
\section*{Conflict of interest}
 The authors declare that they have no conflict of interest.


\bibliographystyle{spphys}       
\bibliography{mybib}  

%
%

\end{document}